\documentclass[preprint,10pt,twocolumn]{revtex4-1}

\newcommand\sss[1][.15cm]{\hspace*{#1}}

\usepackage{mathtools}
\usepackage[a4paper,bindingoffset=0.2in,%
            left=0.5in,right=0.5in,top=1in,bottom=1in,%
            footskip=.25in]{geometry}
\usepackage{blindtext}
\usepackage[utf8]{inputenc}
\usepackage{chngcntr}
\counterwithout{section}{equation}
\usepackage{amsmath}
\usepackage{amsfonts}
\usepackage[toc,page]{appendix}
\usepackage{amssymb}
\usepackage{multirow}
\usepackage{xcolor, soul}
\usepackage{epstopdf}
\usepackage{float}
\usepackage[caption=false]{subfig}
\usepackage{hyperref}
\hypersetup{
     colorlinks   = true,
     citecolor    = red,
     linkcolor    = blue,
     urlcolor     = cyan,
}
\usepackage{ragged2e}
\usepackage{makeidx}
\usepackage[final]{graphicx}
\def\beq{\begin{equation}}
\def\eeq{\end{equation}}
\def\bea{\begin{eqnarray}}
\def\eea{\end{eqnarray}}
\def\nno{\nonumber}
\def\pr{\partial}
\begin{document}
\pagestyle{myheadings}
\title{Probing Non-classicality of Primordial Gravitational Waves and Magnetic Field Through Quantum Poincare Sphere}
\author{Debaprasad Maity}
\email{debu@iitg.ac.in}
\affiliation{Department of Physics, Indian Institute of Technology Guwahati, Assam 781039, India}
\author{Sourav Pal}
\email{pal.sourav@iitg.ac.in}
\affiliation{Department of Physics, Indian Institute of Technology Guwahati, Assam 781039, India}
\date{\today}
\begin{abstract}
The universe is believed to be originated from a quantum state. However, defining measurable quantities for the quantum properties in the present universe has gained interest recently. In this submission, we propose a quantum Poincare sphere as an observable quantity that can hint at the quantumness of primordial gravitational waves and large-scale magnetic fields. The Poincare sphere is defined in terms of quantum stokes operators associated with the polarization of those fields, which can be measured directly. We have further studied the effects of the initial non-BD vacuum on the power spectrum and squeezing parameter of the primordial gravitational waves and magnetic field. We have found that the initial non-BD vacuum increases the value of the squeezing parameter as expected at the end of inflation, which further enhances the possibility of measuring the quantumness of the fields under consideration. To support our results, we further explored the possible Bell violation test for a set of generalized pseudo spin operators defined in the polarization space of those fields.
\end{abstract}
\maketitle
\section{Introduction}
The origin of our universe is believed to be a quantum mechanical phenomenon. 
Any experimental verification of this quantumness would be extremely important to understand the underlying principle of nature.   Therefore, how one can confirm our universe to be of quantum mechanical origin is an important question that has gained significant interest in the recent past. One of the recent endeavors toward understanding this question has been to re-explore the underlying mechanism of the large-scale structure of our universe and look for the observable quantum signatures in it. The most successful inflation \cite{Starobinsky:1980te,Starobinsky:1982ee,Guth:1980zm,Linde:1981mu,Albrecht:1982wi,Linde:1983gd} paradigm, which explains the observed large scale structure of our universe to a great precession, is intrinsically assumed to be quantum mechanical phenomena. Though the observed fluctuations of the matter distribution are successfully explained by quantum fluctuation during inflation, any kind of classical statistical origin of such fluctuation can not be ruled out. Therefore, looking for an unambiguous quantum mechanical signature encoded in the distribution of matter would be a logical step toward establishing the inflationary paradigm itself. Cosmic Microwave Background (CMB) anisotropy,  which directly corresponds to the matter distribution of the universe, is shown to be related to the scalar quantum fluctuations during inflation. Therefore, it has been widely considered as a potential test-bed to look for quantum signature in the early universe \cite{Martin:2012pea,Martin:2015qta,Martin:2016nrr,Martin:2016tbd,Martin:2017zxs}. Inflation is believed to be a fundamental mechanism that produces a primordial gravitational field as well. Hence, further studies have been performed in the context of gravitational waves \cite{Kanno:2019gqw}. The primordial magnetic field is also believed to be originated from inflation \cite{Sharma:2017eps,Sharma:2018kgs,Jain:2012ga,Durrer:2010mq,Kanno:2009ei,Campanelli:2008kh,Ashoorioon:2004rs,Demozzi:2009fu}. Therefore, besides scalar and gravitational waves, the primordial magnetic field can also be an interesting observable which may encode the quantum signature of the early universe. In this paper, we aim to understand this by defining an appropriate observable called quantum Poincare sphere. We will consider both gravitational and electromagnetic fields in a unified framework for our present discussion.  
 
 Measuring the gravitational waves and cosmic magnetic field is a long-standing endeavor with resounding successes. Detection of gravitational waves over the last few years by LIGO and Virgo \cite{Abbott:2016blz,lv2,lv3,lv4,lv5,lv6,lv7,lv8,lv9,lv10,lv11,lv12,lv13,lv14} has opened up a new era that has the potential to answer all these fundamental questions about the origin of our universe in the near future. Magnetic fields of order a few micro-Gauss ($\mu $G) with coherence length of hundred kilo-parsecs (Kpc) scale \cite{Grasso:2000wj,Kronberg:1993vk,Widrow:2002ud} has been observed in the galaxies and galaxy clusters. 
 The intergalactic void may also host a weak $\sim 10^{-16}$ Gauss magnetic field, with the coherence length as large as Mpc scales \cite{Durrer:2013pga}. 
The magnetic field of Mpc coherence scale is difficult to explain by any known classical processes. Further, at the astrophysical scale, to generate a magnetic field of  $\mu $G order, one needs a very weak seed magnetic field. To explain the  fundamental origin of such magnetic field at different scales, inflationary magnetogenesis \cite{Kobayashi:2014sga,Kobayashi:2019uqs, Ferreira:2013sqa,Martin:2007ue,Ratra:1991bn} is believed to be a natural mechanism. Similar to scalar and gravitational perturbation, the quantum electromagnetic fluctuations during inflation can be amplified to produce a large-scale magnetic field. Apart from being present in the void, such a magnetic field can then act as a seed at small scales and get enhanced to the galactic scale of $\mu$G  order by the well-known \textit{Galactic dynamo} mechanism\cite{Kronberg:1993vk,dolginov,asseo}.
 
Motivated by these facts, our primary goal of this letter would be to construct unique observables for both gravitational waves and magnetic fields, which can in principle be measured in the present universe and provide important quantum signatures. Our analysis will be along the line of the squeezed state formalism \cite{Martin:2012pea,Martin:2015qta,Martin:2016nrr}. Furthermore, by constructing appropriate observables measured in this state, we will solidify our proposal for quantum signatures.  

\section{Magnetogenesis in squeezed state formalism}
\begin{figure}
\begin{center}
\includegraphics[scale=0.7]{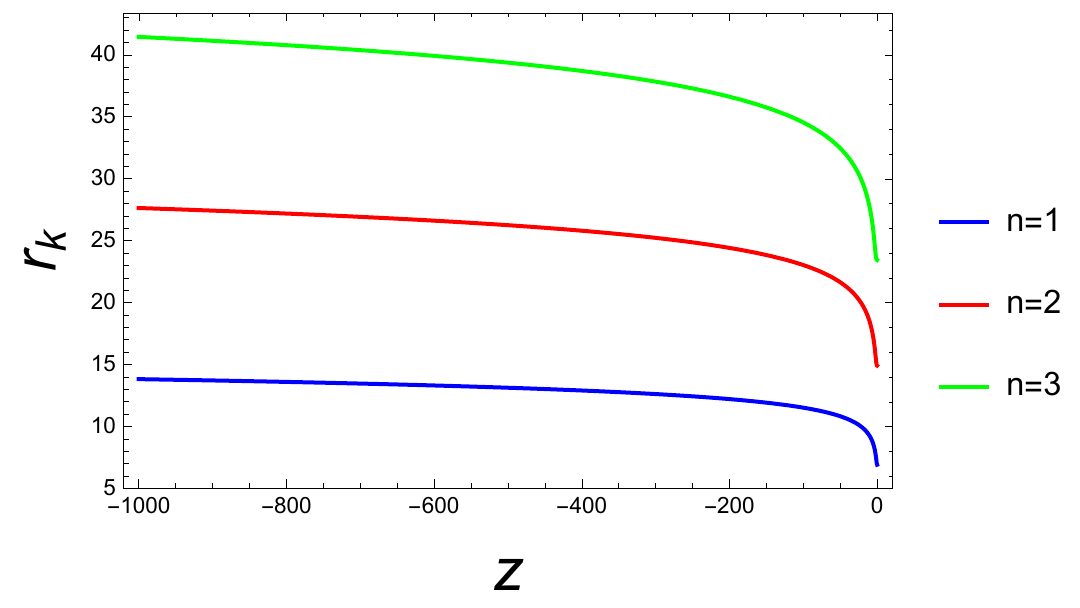}
\caption{Evolution of the squeezing parameter $r_k$ with $z=k\tau$ for different values of the coupling parameter $n$}\label{evolution_r}
\end{center}
\end{figure}

We begin with detailed account of the primordial magnetogenesis in the squeezed state formalism. The conformal symmetry broken electromagnetic (EM) Lagrangian is taken as \cite{Ratra:1991bn}, 
\begin{equation}
\label{action}
S=-\frac{1}{4}\int d^4x \sqrt{-g}I^2(\tau)F_{\mu \nu}F^{\mu \nu},
\end{equation}
where, $F_{\mu \nu} =\partial_{\mu}A_{\nu}-\partial_{\nu}A_{\mu}$ is EM field strength tensor with $A_\mu$ being the vector potential. $I(\tau)$ is arbitrary EM coupling which breaks the conformal invariance of the EM field. Generally in case of inflationary magnetogenesis this coupling is taken as a function of the inflaton field, which goes to 1 at the end of inflation. Following the literature it can be written as
\bea\label{coupling}
    I(\tau)= 
\begin{cases}
    \left(\frac{a_{end}}{a}\right)^{2n},& \text{if } \tau < \tau_{end}\\
    1,              & \tau\geq \tau_{end}
\end{cases}
\eea
Where $a$ is the scale factor at any arbitrary time $\tau$, $a_{end}$ is the scale factor at the end of inflation, and $n$ is the coupling parameter.
 The background is taken to be spatially flat FLRW metric with scale factor $a(\tau)$, 
\bea\label{metric}
ds^2 = a(\tau)^2( -d\tau^2 + dx^2 + dy^2 + dz^2),
\eea
Because of spatial flatness, components of the vector potential can be defined in terms of irreducible components as
$
A_\mu=\left(A_0,\partial_iS+V_i\right)$ with the traceless condition $ \delta^{ij}\partial_iV_j=0.$
In terms of these components the action Eq.\eqref{action} becomes,
\begin{equation}
\label{action_pot}
S=\frac{1}{2}\int d\tau d^3xI(\tau)(V_i^\prime {V^i}^\prime-\partial_iV_j\partial^i V^j).
\end{equation}
It is important to note that the action does not depend on the scale factor explicitly because of the inherent conformal invariance of free electrodynamics. The spatial index will be raised or lowered by the usual Kronecker delta function. The Fourier expansion of $V_i$ is taken as 
\begin{equation}
\label{mode_EM}
V_i(\tau,x)=\sum_{p=1,2}\int\frac{d^3k}{(2\pi)^3}\epsilon^{(p)}_i(\textbf{k})  e^{i\textbf{k.x}}u_k^{(p)}(\tau),
\end{equation}
where the reality condition of the field implies $u^{p}_{-k} = u^{p*}_k$. Here, $\epsilon_i^{(p)}(k)$ is the polarization vectors corresponding to two modes $p=1,2$, which satisfy  
$
\epsilon_i^{(p)}(\textbf{k}) k_i=0$ and $\epsilon_i^{(p)}(\textbf{k})\epsilon_i^{(q)}(\textbf{k})=\delta_{pq}.
$
Dynamics of the field will be governed by following mode function equation,
\beq
\label{eq10}
u^{(p)\prime\prime}_k+\frac{I^\prime}{I}u^{(p)\prime}_k+k^2u^{(p)}_k=0 
\eeq
Where the prime denotes derivative with respect to the proper time $\tau$. In terms of $u^k_{p}$ and its associated conjugate momentum $\pi^{p}_k = I {u^{p}_{k}}'$, the Hamiltonian becomes,
\bea
{\cal H} = \frac 1 2\sum_{p=1,2}\int\frac{d^3k}{(2\pi)^3}\left( \frac{\pi^{p}_k\pi^{p*}_k}{I} +  I k^2 u^k_{p*}u^k_{p} \right). 
\eea
Treating the above canonically conjugate variable as operators, we express those in terms of creation and annihilation operators, 
\bea
\pi^p_k = -i \sqrt{\frac {k}{2}} (a^p_k - a_{-k}^{\dagger(p)})~;~ u^p_k = \sqrt{\frac {1}{2k}} (a^p_k + a_{-k}^{\dagger(p)}),
\eea
with $[a_k^{(p)},a_h^{\dagger(q)}]=(2\pi )^3\delta^{pq}\delta^{(3)}(\textbf{k-h})$ being the fundamental commutation relation. In terms of the operators, Hamiltonian of each mode $k$ becomes,
\begin{eqnarray}
\label{eq7}
H_k =k\sum_p \left(\frac{{\cal I}_1}{2}\left(a_k^{p\dagger} a^p_k 
+ a^p_{-k} a_{-k}^{p\dagger}\right)
+\frac{{\cal I}_2}{2}\left(a^p_k a^p_{-k} + a^{p\dagger}_{-k}a^{p \dagger}_k\right)\right)\nonumber
\end{eqnarray}
where, ${\cal I}_1 = (I+1/I),~{\cal I}_2 = (I-1/I)$. To incorporate the time evolution of the creation and annihilation operators we employ Bogoliubov transformation as, $
a^p_k(\tau)=\alpha_k(\tau)a^p_k(0)+\beta_k(\tau) a^{p\dagger}_{-k}(0)$,
where, Bogoliubov coefficients satisfy the relation $
\vert\alpha_k\vert^2-\vert\beta_k\vert^2=1 $, which is parametrized by,
\begin{equation}
\label{eq11}
\alpha_k=e^{i\theta_k} \cosh(r_k) ~;~ \beta_k= - e^{-i\theta_k+2i\phi_k}~ \sinh(r_k) 
\end{equation}
$(r_k,\phi_k)$ symbolize the squeezing parameter and squeezing angle of the quantum state respectively. They satisfy
\bea\label{PMF_squ_evol}
\phi_k' &=& -\frac {k {\cal I}_1}{2} + \frac {k {\cal I}_2}{2} \cos 2\phi_k ~\coth 2r_k, \\
r_k' &=& \frac{k{\cal I}_2}{2}  \sin2\phi_k ~;~
\theta_k' = -\frac{k {\cal I}_1}{2} + \frac{k {\cal I}_2}{4} \cos2\phi_k \tanh r_k . \nonumber
\eea
If one chooses the Bunch-Davis vacuum with the coupling function mentioned in Eq.\eqref{coupling} then, the solution of those parameters can be expressed as
\bea \label{squeeze-param}
r_k &=& \sinh^{-1} |\beta_k| ~;~~\phi_k = \frac{1}{2} Arg(\alpha_k \beta^*_k)\nonumber
\eea
If we calculate the Bogoliubov coefficients with the Bunch Davies condition, it turns out as 
\bea\label{Bogo_PMF}
\beta_k^* &=& \left(\frac{\pi z}{8 I}\right)^{\frac12}\left(H^1_{-n+1/2}(z) + i I  H^1_{-n-1/2}(z) \right)
\nno\\
\alpha_k &=& \left(\frac{\pi z}{8  I}\right)^{\frac12}\left(H^1_{-n+1/2}(z) - i I H^1_{-n-1/2}(z) \right)
\eea
 Where $z=k\tau$. In this parametrization we have assumed that the Hubble constant (H) remains constant throughout the inflationary era. And it is taken as $H_{end}$. We can find out the time evolution of the squeezing parameter $(r_k)$ from Eqs.\eqref{squeeze-param}. The evolution of $r_k$ with $z=k\tau$ is shown in Fig.\ref{evolution_r} for different values of coupling parameter $n$.

To this end, let us give a connection formula with the observable quantities such as electric and magnetic correlation function with the aforementioned squeezing parameters as
\bea
<E_{\mu}(x) E^{\mu}(y)> = \int \frac {d^3 k}{4 \pi k^3} e^{i k\cdot(x-y)} \mathcal{P}_{E}(\tau,k) \nonumber \\
 <B_{\mu}(x) B^{\mu}(y)> = \int \frac {d^3 k}{4 \pi k^3} e^{i k\cdot(x-y)} \mathcal{P}_{B}(\tau,k) 
 \eea
Where, electric and magnetic power spectrum defined in momentum space turned out to be 
\begin{eqnarray}\label{power_spectra}
\mathcal{P}_{E}(\tau,k) &=& \frac {k^4}{4\pi^2 a^4 I^2} \sum_{p=1,2} \vert\alpha_k^{(p)}-\beta_k^{(p)}\vert^2\nno\\
&=& \frac {k^4}{4\pi^2 a^4 I^2} \sum_{p=1,2} (\cosh 2r_k^p - \sinh 2r_k \cos 2 \phi_k^p ) \nno\\
\mathcal{P}_{B}(\tau,k)& =& \frac {k^4}{4\pi^2 a^4 I^2} \sum_{p=1,2}\vert\alpha_k^{(p)}+\beta_k^{(p)}\vert^2\\
&=& \frac {k^4}{4\pi^2 a^4 I^2} \sum_{p=1,2}  (\cosh 2r_k^p + \sinh 2r_k^p \cos 2 \phi_k^p ) \nno
\end{eqnarray}
 After the end of inflation, the conformal invariance is restored, and thus there is no further production of the EM field. If we consider the instantaneous reheating scenario, the universe readily transforms into a plasma state. Which makes the electric field vanish due to the high conductivity of the medium. This also freezes the magnetic field produced in the inflationary era. As there is no production of magnetic field in the post-inflationary era, the magnetic field power spectrum decays as $a^{-4}$.

\begin{figure}\label{aH_vs_a}
\centering
\includegraphics[scale=0.82]{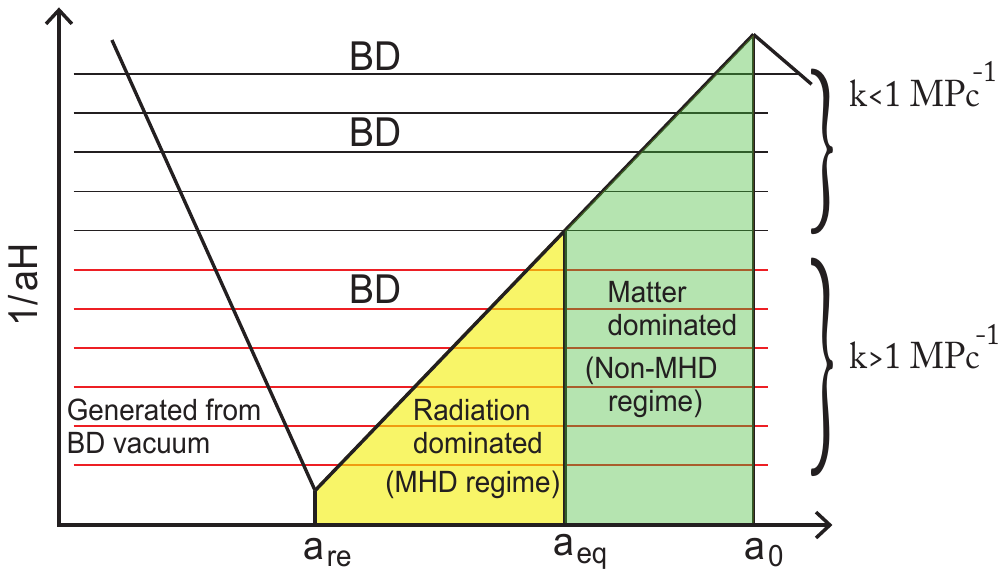}
\caption{$1/aH$ is plotted against $a$ here. The region in yellow shows the radiation dominated era where the MHD takes over the evolution of the magnetic field, and the modes re-entering the horizon in this era lose the information about initial quantum production. The green shaded region represents the matter dominated era, and the modes re-entering the horizon in this era will not undergo MHD processes. BD stands for Bunch-Davis vacuum.}
\end{figure}

At the present time, the magnetic power spectrum turns out to take the following form,
\bea
\mathcal{P}_{B0}\propto \left(\frac{k}{a_0}\right)^{6-2n}e^{2(n-1)N}~~
\eea
Where $(N,a_0)$ are the inflationary e-folding number and the present value of the scale factor, respectively. By this mechanism, therefore, a large scale magnetic field of order $10^{-15}-10^{-22}$ Gauss can be generated on Mpc scales \cite{Martin:2007ue,Vachaspati:2016xji}, once an inflation model is considered. The magnetic field generated during the inflationary era is produced from quantum fluctuations. Subsequently, those modes will evolve throughout the entire cosmological evolution. As mentioned earlier, due to the large conductivity electric field vanishes, but the magnetic field freezes after the end of inflation. However, the subsequent dynamics of this frozen magnetic field will depend on the individual modes. 
The	magnetic modes which re-enter the Hubble horizon during the radiation era interact with the plasma and further evolve through magneto-hydrodynamic (MHD) evolution. Due to this MHD effect, the information of the inflationary origin of the magnetic field from the Bunch-Davis vacuum will be erased for those magnetic modes. On the other hand, the modes with larger wavelengths will re-enter the horizon during matter dominated era and will not interact with the plasma. Hence, no MHD evolution will take place. Therefore, those large-scale magnetic fields are expected to preserve the information of their quantum origin during inflation. The above mentioned fact is schematically illustrated in Fig.\ref{aH_vs_a}. In the following discussion, we estimate the approximate scale above which information of the quantum origin of the magnetic field will be preserved. We calculate the wavenumber of the mode which re-enters the horizon during matter-radiation equality, where following relation is satisfied,
\beq
\frac{T_{eq}}{T_{re}}=\frac{a_{re}}{a_{eq}},
\eeq
where $a_{eq}$ is the scale factor at the matter-radiation equality, and $H_{eq}$ is the Hubble parameter at that point. The temperature at that point is $T_{eq}\approx 10^{-9}$ GeV, $T_{re}$ is the reheating temperature. Considering the instantaneous reheating scenario, it is generically observed that $T_{re}\approx  10^{15}$ GeV. Furthermore, the entropy conservation leads to the following relation,  \bea\label{entropy}
a_{eq} H_{eq}=a_{re} H_{re} \frac{a_{re}}{a_{eq}} \implies
k_{eq} = 10^{-24} k_{re},
 \eea 
where $k_{eq}$ and $k_{re}$ are the modes associated with the radiation-matter equality and the end of reheating. For instantaneous reheating, $k_{re} \sim 10^{-15}$ GeV, which immediately gives $k_{eq} \sim 10^{-39}~ \mbox{GeV} \sim 1~ \mbox{Mpc}^{-1}$.
From this estimation, one observes that any mode  $k < k_{eq} \sim ~1 ~\mbox{Mpc}^{-1}$ re-enters the horizon during the matter-dominated era, is expected to carry information about its quantum origin, as those are not affected by causal MHD processes. A schematic of the modes that re-enters the horizon during the different era is given in Fig.\ref{aH_vs_a}.
\\

Therefore, from the observation of magnetic field strength and structure for large-scale modes, we may be able to decode the quantum nature early universe through observables such as squeezing parameters, which encode the evolution of the field. 

\section{Primordial Gravitational waves in squeezed state formalism}
\begin{figure}[t]
\begin{center}
\includegraphics[scale=0.6]{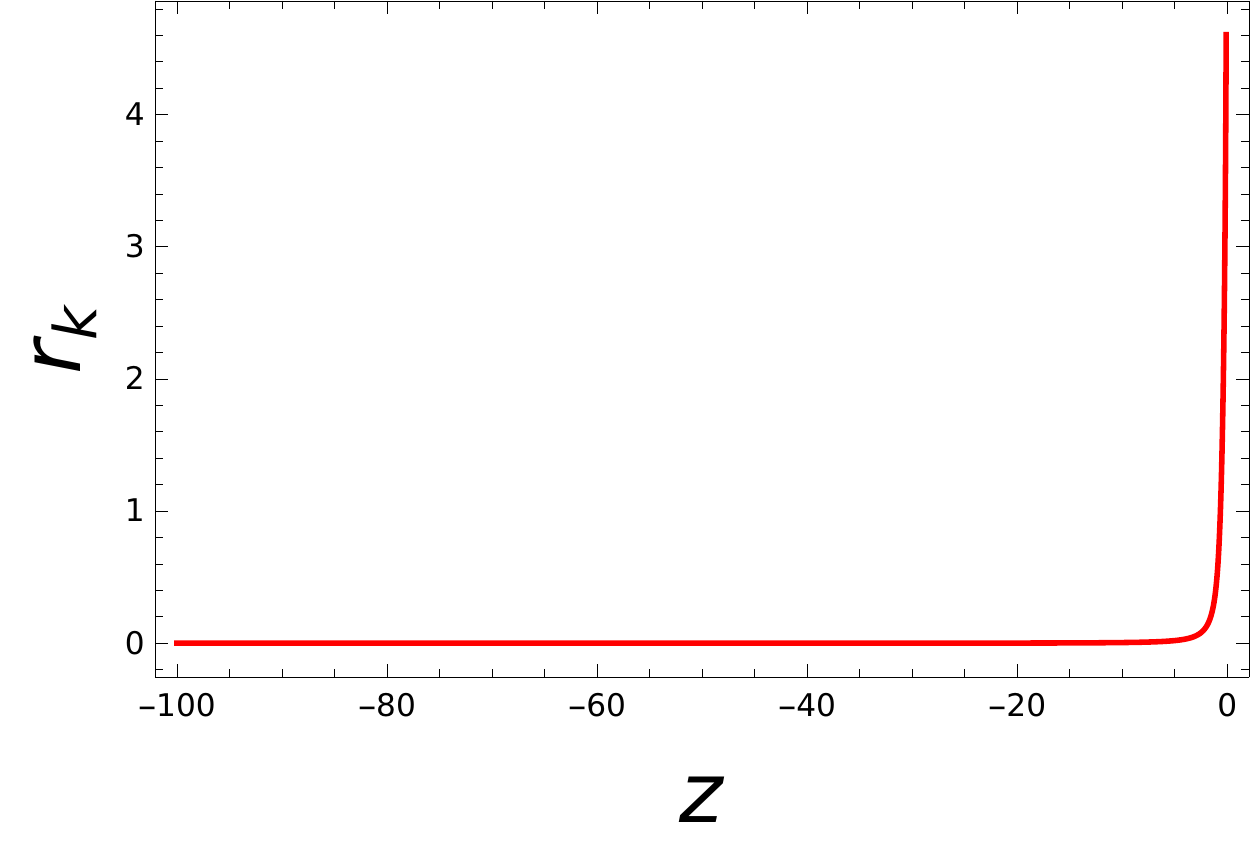}
\caption{Evolution of the squeezing parameter $r_k$ with $z=k\tau$ in case of gravitational waves}\label{grav_evolution}
\end{center}
\end{figure}

Primordial gravitational wave (GW) has been studied already in the squeezed state formalism \cite{Kanno:2019gqw,Kanno:2020lpi,Koh:2004ez}. Hence we will quote the main results here. The formalism would be the same as discussed for the electromagnetic case. The physical component of the  gravitational waves is defined by the transverse traceless part of the metric perturbation, 
\beq\label{PGW_metric}
ds^2= a^2(\tau)\left[-d\tau^2+(\delta_{ij}+h_{ij})dx^i dx^j\right]
\eeq
The tensor perturbation here $h_{ij}$ satisfies the relation $\pr_i h^{ij} = 0 ,~h_i^i=0 ~\mbox{and}~ |h_{ij}|\ll \delta_{ij}$. 
\beq\label{mode_expansion}
h_{ij}(x,\tau)= \frac{1}{a}\int \frac{d^3 k}{(2\pi)^3}e^p_{ij}~h^p_k ~e^{i k\cdot x}
\eeq
Where $e_{ij}$ is the polarization tensor for the GW and it satisfies the relation $e^{m *}_{ij}e_{ij}^n = 2\delta^{mn}$. Here $p$  is the polarization index corresponding to the GW. Like the electromagnetic fields, $p$ assumes two values corresponding to the two polarization modes. The conjugate momentum corresponding to $h_k$ is
\beq
\pi_k^{(p)}=h_k^{'(p)}-\frac{a'}{a} h_k^{(p)}
\eeq
With this in hand the Hamiltonian turns out as
\bea
\mathcal{H}&=&\frac{1}{2}\sum_{p=1,2}\int\frac{d^3 k}{(2\pi)^3}~k\left(a_k^{(p)}a_k^{\dagger(p)}+a_{-k}^{\dagger(p)}a_{-k}^{(p)}\right)\nno\\
&+&i\frac{a'}{a}\left(a_k^{\dagger(p)}a_{-k}^{\dagger(p)}-a_k^{(p)}a_{-k}^{(p)}\right)
\eea
Now as in the case of electromagnetic case we can calculate the Bogoliubov coefficients as
\bea\label{Bogo_PGW}
\alpha_k=\left(1+\frac{i}{k\tau}-\frac{1}{2 k^2 \tau^2}\right)~;~\beta_k =\frac{1}{2 k^2 \tau^2}
\eea
 Here also, the Bogoliubov coefficients are represented in terms of the squeezing parameter $(r_k)$ and the squeezing angle $(\phi_k)$ as in Eq.\eqref{eq11}. 
The evolution of the squeezing parameter and the squeezing angle in the case of primordial gravitational waves is dictated by the following differential equations:
 \bea\label{evol_grav}
 \phi'_k &=&  -k + \frac{a'}{a} \sin (2\phi_k) \coth(2 r_k)\\
r'_k &=& -\frac{a'}{a} \cos(2\phi_k)~;~\theta'_k = -k +\frac{a'}{a}\sin (2\phi_k)\tanh (r_k)\nno 
 \eea 
 Moreover, the evolution of the squeezing parameter with the conformal time turns out, as shown in Fig.\ref{grav_evolution}.
\\
And the GW power spectra in terms of the squeezing parameter $(r_k)$ and squeezing angle $(\phi_k)$ turns out as
\bea\label{PGW_power_spectra_present}
\mathcal{P}^{GW}(\tau,k)&=& \frac{2k^2}{\pi^2 M_{Pl}^2 a^2}\sum_{p=1,2} |\alpha_k^{(p)}+\beta_k^{(p)}|^2\nno\\
&=&\frac{2k^2}{\pi^2 M_{Pl}^2 a^2}\sum_{p=1,2}\bigg(\cosh(2r_k^{(p)})\nno\\
&+& \sinh(2r_k^{(p)}) \sin (2\phi_k^{(p)})\bigg)
\eea
Where $M_{pl}=(8\pi G)^{-1/2}$ is the reduced Planck mass. From the above expression, it is evident that the power spectra for GW also depend on the squeezing parameter $r_k$, which can be measured from the strengths. Towards this end, let us point out an important observation regarding the distinct decreasing behavior of the squeezing parameter for gauge field (see Fig.\ref{evolution_r}) from that of the gravitational wave (see Fig.\ref{grav_evolution}). This distinct behavior of $r_k$ for gauge field originates from underlying conformal invariance. Because of this property, the evolution of the squeezing parameter of the gauge field does not depend on the cosmological scale factor as the background is conformally flat. Hence, it depends only on the gauge coupling parameter $I(\tau)$. The coupling function Eq.\ref{coupling} is assumed in such a way that it decreases and becomes unity at the end of inflation which restores the standard conformal invariant Maxwell's theory. As a result $r_k = \sinh^{-1}|\beta_k|$ also decrease with time. After the end of inflation, the conformal invariance of the EM field is restored, and there will be no more evolution of the squeezing parameter $r_k$ from background expansion. In the asymptotic $k\eta \rightarrow 0$ limit the values of the squeezing parameter turned out to be $r_k$= 6.9, 14.9 and 23.4 for $n$= 1, 2, 3 respectively. However, the dynamics of $r_k$ for gravitational waves explicitly depend on the scale factor, which leads $r_k$  to increase in time, as depicted in Fig.\ref{grav_evolution}. Therefore, for both the cases the information about the quantum origin of the fields should be encoded into the non-zero value of their respective squeezing parameters.
\section{Modifications due to initial Non Bunch Davies Vacuum}
 When we solved for the mode functions of gauge fields and the tensor perturbations, in the infinite past, we considered the usual Bunch Davies (BD) vacuum, parametrized by Bogolubov parameter values $\alpha_k=1$ and $\beta_k=0$ for $k\tau\rightarrow -\infty$. However, generalizing the initial condition to non-BD vacuum has been studied extensively for scalar, and tensor perturbations in the literature \cite{Holman:2007na, Ashoorioon:2010xg, Ashoorioon:2013eia, Meerburg:2009ys, Meerburg:2009fi,Ashoorioon:2012kh}. Such studies for gauge fields have not been considered before, and we take this opportunity to investigate such effects of non-BD vacuum on the Primordial magnetic field. For the sake of our present study, we also discuss non-BD modification of primordial gravitational waves power spectrum. 
 \subsection{Modifications to primordial magnetic field power spectrum due to non-BD vacuum}
 The time-dependent magnetic field power spectrum can be expressed written in terms of the Bogoliubov coefficients as \cite{Kobayashi:2019uqs}
 \beq
 \mathcal{P}_B(k,\tau)=\frac{k^4}{2\pi^2 a^4 I^2}\sum_{p=1,2}|\alpha_k^p+\beta_k^p|^2\nno .
 \eeq
 In the past infinity, if we take the BD initial condition, i.e., $\alpha_k=1, \beta_k=0 $, the power spectrum will turn out as
 \beq
\mathcal{P}_B(k)\vert_{k\tau\rightarrow -\infty}=\frac{k^4}{\pi^2 a^4 I^2} .
 \eeq
Usually, the non-BD initial state is parametrized by non-vanishing $\beta_k$ for each mode, and the power spectrum at $k\tau \rightarrow \infty$ will be,
 \bea\label{spec_PMF_B}
 \mathcal{P}_B(k)\vert_{k \tau\rightarrow -\infty}=\frac{k^4}{2\pi^2 a^4 I^2}\sum_{p=1,2}|\alpha_k^{0(p)}+\beta_k^{0(p)})|^2 
 \eea
 Where $\alpha_k^{0(p)}$ and $\beta_k^{0(p)}$ represents the Bogoliubov coefficients in the limit $k\tau\rightarrow -\infty$. For numerical purpose we define a dimensionless ratio $\delta_k^{(p)}={\beta_k^{0(p)}}/{\alpha_k^{0(p)}}$. For BD vacuum, this ratio goes to zero. The Bogoliubov coefficients follow the relation
\beq\label{eq26}
|\alpha_k|^2-|\beta_k|^2=1
\eeq
In the infinite past also, the Bogoliubov coefficients follow the relation in Eq.\eqref{eq26}. Thus the quantity $|\alpha_k^{0(p)}|^2$ can be written in terms of $\delta_k^{(p)}$ as
\bea
|\alpha_k^{0(p)}|^2-|\beta_k^{0(p)}|^2&=&1\nno\\
|\alpha_k^{0(p)}|^2\left(1-\frac{|\beta_k^{0(p)}|^2}{|\alpha_k^{0(p)}|^2}\right)&=&1\nno\\
|\alpha_k^{0(p)}|^2\left(1-|\delta_k^p|^2 \right)&=& 1\nno
\eea
Finally, it turns out as
\beq\label{alpha_k}
|\alpha_k^{0(p)}|^2 = \frac{1}{1-|\delta_k^{(p)}|^2}
\eeq
Now using Eq.\eqref{alpha_k} the power spectrum can be written in terms of the quantity $\delta_k^{(p)}$ as,
\bea
 \mathcal{P}_B(k)\vert_{k\tau \rightarrow -\infty}&=&\frac{k^4}{2\pi^2 a^4 I^2}\sum_{p=1,2}|\alpha_k^{0(p)}|^2 |1+\delta_k^{(p)}|^2\nno\\&=&\frac{k^4}{2\pi^2 a^4 I^2}\sum_{p=1,2}\frac{|1+ \delta_k^{(p)}|^2}{1-|\delta_k^{(p)}|^2}\nno\\
 &=& \frac{k^4}{2\pi^2 a^4 I^2}\sum_{p=1,2}(1+g_k^{(p)}).
 \eea
Where $g_k^{(p)}$ encodes the information of the non-BD initial excited state. Here the factor $g_k^{(p)}$ is related to $\delta_k^{(p)}$ as 
\beq\label{delta_g}
\delta_k^{(p)}=\frac{g_k^{(p)}}{2+g_k^{(p)}} .
\eeq
And the Bogoliubov coefficients can be written in terms of $g_k^{(p)}$ as
\beq\label{bogo_exc}
\alpha_k^{0(p)}=\frac{2+g_k^{(p)}}{2\sqrt{1+g_k^{(p)}}}~~;~~\beta_k^{0(p)}=\frac{g_k^{(p)}}{2\sqrt{1+g_k^{(p)}}} .
\eeq
For the simplicity of our calculation, we have taken that there is no relative phase difference between the two Bogoliubov coefficients in the past. It is evident from Eq.\eqref{bogo_exc} that the values of these coefficients is fixed by particular choice of the function $g_k^{(p)}$.
\subsection{Modifications of primordial gravitational wave power spectrum due to Non-BD vacuum}
As we discussed in the previous section for primordial magnetic fields,  the gravitational waves will also get similar modifications non-BD vacuum state. The power spectrum for the PGW can be written in terms of the Bogoliubov coefficients as
\beq
\mathcal{P}^{GW}(k,\tau)=\frac{2k^2}{\pi^2 M_{pl}^2 a^2}\sum_{p=1,2}|\alpha_k^{(p)}+\beta_k^{(p)}|^2
\eeq
For BD initial condition, the power spectrum the in the infinite past looks like
\beq
\mathcal{P}^{GW}(k)\vert_{k\tau\rightarrow -\infty}=\frac{4k^2}{\pi^2 M_{pl}^2 a^2}
\eeq
Now taking into consideration initial excited state, exactly like in case of the PMF we can write it in terms of the ratio $\delta_k^{(p)}={\beta_k^{0(p)}}/{\alpha_k^{0(p)}}$.  Following the steps from Eq.\eqref{eq26} to Eq.\eqref{alpha_k} we can obtain the power spectrum as
\bea
\mathcal{P}^{GW}(k)\vert_{k\tau\rightarrow -\infty}&=&\frac{2k^2}{\pi^2 M_{pl}^2 a^2}\sum_{p=1,2}\frac{|1+\delta_k^{(p)}|^2}{1-|\delta_k^{(p)}|^2}\nno\\
&=& \frac{2k^2}{\pi^2 M_{pl}^2 a^2}\sum_{p=1,2}(1+g_k^{(p)})
\eea
The factor $g_k^{(p)}$ is related to $\delta_k^{(p)}$ as described in Eq.\eqref{delta_g}.
And the Bogoliubov coefficients are related to the factor $g_k^{(p)}$ as described in Eq.\eqref{bogo_exc}.
The equations \eqref{Bogo_PMF} and \eqref{Bogo_PGW} shows that for BD initial state one get $\beta_k=0$ in the asymptotic past. However, for the non-BD initial state, $\beta_k$ is non-vanishing. In our squeezing state formalism, this non-zero $\beta_k$ in the infinite past suggests that 
the initial is squeezed.\\ 
In the following, we consider a particular model to show the deviation from the regular BD vacuum. We assume the form of $g_k^{(p)}$ to be a lognormal distribution following from the reference \cite{Ragavendra:2020vud}.
\beq\label{beta_mod}
g_k^{(p)}=\frac{\gamma}{\sqrt{2\pi (\Delta_k^{(p)})^2}}\exp\left[-\frac{ln^2(k/k_i)}{2(\Delta_k^{(p)})^2}\right] .
\eeq
Where $\gamma$ represents the strength of the distribution, $\Delta_k$ denotes the width of the Gaussian, and $k_i$ denotes the location of the peak of the distribution. If we set $\gamma=0$, the non-BD states boil down to the standard Bunch-Davies vacuum state. Furthermore, for the modes  $k>>k_i$, $g_k^{(p)}\rightarrow 0$, which implies that the small wavelength modes perceive BD vacuum. For example, if we choose a particular value of the parameters of the non-BD state, $k_i=1~\mbox{MPc}^{-1}$, $k=0.8~\mbox{MPc}^{-1}$, $\gamma=4.5$ and $\Delta_k^p$ as unity for both the polarization, then the Bogoliubov coefficients turns out to be $\alpha_k= 1.13,~\beta_k= 0.53$ for both the cases. These new values also satisfies the relation $|\alpha_k|^2-|\beta_k|^2=1$.
For the case of BD vacuum, the initial value of the squeezing parameter $r_k$ is zero. However, from Eq.\eqref{squeeze-param}, we can calculate the value of the squeezing parameter at the beginning of the evolution, and the value turns out as $r_0=0.51$ for the set of the parameters mentioned above. The effect of this initial excited state can be realized from the evolution of the squeezing parameter $(r_k)$  with $r_0$ as the initial condition. It can in principle, be observed in the present day. 
 \begin{figure}
\begin{center}
\includegraphics[scale=0.87]{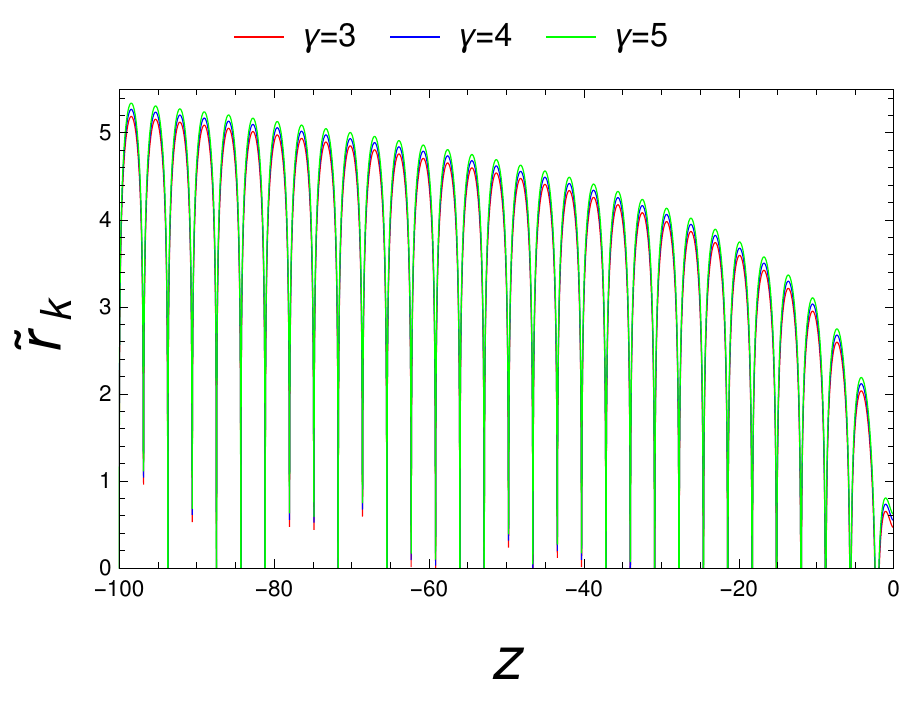}
\caption{Here the evolution of the squeezing parameter $(r_k)$ of the primordial magnetic field for coupling parameter $n=1$ is plotted against $k\tau$. On the y-axis we have plotted $(r_k-9)$ (for scaling) and on the x-axis $z=k\tau$ is plotted. The three colors correspond to different values of the initial squeezing parameter $r_0$, the variation in $r_0$ is obtained by varying the strength of the distribution $\gamma$ introduced in Eq.\eqref{beta_mod}, while keeping the value of $k$ fixed at $0.8~MPc^{-1}$.}
\end{center}\label{evol_exc}
\end{figure}
As the $r_k$ is non-zero initially, we numerically solve Eq.\eqref{PMF_squ_evol} and study the evolution of the squeezing parameter for magnetic field $r_k$ with $z=k\tau$ as shown in Fig.\ref{evol_exc}. In the case of primordial gravitational waves, the squeezing parameter asymptotically goes to infinity as $\tau\rightarrow 0$, like the BD vacuum. However, we can always calculate the value at finite but small $\tau$, which can be connected with the present-day observation. For a better understanding of the modifications due to initial excitation, we tabulate the values of the squeezing parameter at the end of inflation for the case of primordial magnetic fields. Here we only show the changes in the final values $r_k$ (see Tab.\ref{tab1}) due to initial squeezing $r_0$ for the coupling parameter value $n=1$.
\begin{table}
\begin{tabular}{|c|c|c|c|c|c|c|c|c|}
\hline
\multicolumn{3}{|c|}{Bunch Davies case} & \multicolumn{4}{c|}{Non Bunch Davies case}\\
\hline
Initial&  Initial & Final & \multirow{2}{*}{Parameters} &Initial&  Initial & Final\\
value & value & value&&value & value & value\\
of $\beta_k$ & of $r_k$ & of $r_k$& ($\gamma, k ~\text{in}~ \scriptsize{MPc^{-1}}$)&of $\beta_k$ & of $r_k$ & of $r_k$ \\
\hline
\multirow{6}{*}{0} & \multirow{6}{*}{0} & \multirow{6}{*}{6.9}  & $\gamma= 3,~k =0.8$ & 0.396 & 0.387 & 9.468\\\cline{4-7}
&&&$\gamma= 4,~k =0.8$ & 0.487 & 0.469 & 9.550\\\cline{4-7}
&&&$\gamma= 5,~k =0.8$ & 0.567 & 0.540 & 9.621\\\cline{4-7}
&&&$\gamma= 4,~k =0.5$ & 0.418 & 0.406 & 9.487\\\cline{4-7}
&&&$\gamma= 4,~k =0.7$ & 0.474 & 0.0.458 & 9.539\\\cline{4-7}
&&&$\gamma= 4,~k =0.9$ & 0.493 & 0.475 & 9.556\\
\hline
\end{tabular}
\caption{The final values of the squeezing parameter in the case of the primordial magnetic field are listed due to different $r_0$. The parameters $\gamma$ and $k$ are the variables of the initial value of squeezing parameter $r_0$. The final value denotes the value of the squeezing parameter at the end of inflation.}\label{tab1}
\end{table}
To this end, let us briefly discuss the physical interpretation of the excited state. The initial excited state can be regarded as a Bogoliubov transformation of the non-excited BD vacuum state. If $u_k$ is the mode function corresponding to BD vacuum, then if we do a transformation as
\beq
\tilde{u}_k= \alpha_k u_k+ \beta_k u_k^*\nno
\eeq
Then the mode $\tilde{u}_k$ is the mode function of the excited state containing $n_k= |\beta_k|^2$ number of particles with momentum $k$. 

\section{Squeezed quantum states}
In order to perform experiments for quantum signatures, following \cite{Martin:2016tbd}, we  construct the quantum state generalizing for two polarization states of the photon and graviton. For each polarization mode we already defined associated squeezing parameters, $(r_k^{1,2}, \phi_k^{1,2})$. A generic quadratic Hamiltonian can always be expressed in terms of those parameters and associated so called squeezing operator $S\left(r_k,\phi_k\right)=
e^{B_k}$ and rotation operator $R_k = e^{D_k}$ for each individual polarization mode, with \cite{Martin:2012pea,Martin:2015qta}
\bea
B_k &=& r_k e^{-2i\phi_k} a_{-k} a_k - r_k e^{2i\phi_k} a^{\dagger}_{-k} a^{\dagger}_{-k}\nonumber \\
D_k &=& -i \theta_k a^{\dagger}_{-k} a_k - i\theta_k a^{\dagger}_{-k} a_{-k} 
\eea
These are unitary operators  
which can be shown to form the  unitary time evolution operator for the quantum state \cite{Martin:2016tbd}. During inflation, it is generically assumed that all the quantum state start to evolve from the Bunch-Davies vacuum. Because of the coupling of the fields under study with the classical background, the vacuum state will evolve to squeezed state, which can be expressed as follows, 
\begin{eqnarray}
\label{state_squeeze}
&&\vert\psi\rangle =S_2(r_k^2,\phi_k^2) R_2(\theta_k^2) \vert0\rangle \otimes  S_1(r_k^1,\phi_k^1)R_1(\theta_k^1)\vert0\rangle \\
&=& \frac{1}{{\cal A}}\sum_{n,m=0}^\infty e^{-2i(n\phi_k^1 + m\phi_k^2)}\tanh^n r_k^1\tanh^m r_k^2\vert n_{k,-k}^1,m_{k,-k}^2 \rangle .\nonumber 
\end{eqnarray}
Where, ${\cal A} = \cosh (r_k^1) \cosh(r_k^2)$.
The position space representation of  Eq.\eqref{state_squeeze} assume the following form, 
\begin{equation}
\label{state}
\langle q^{1,2}_A,q^{1,2}_B \vert\Psi\rangle=\frac{1}{\pi} \prod_p \frac{e^{(X_p {q_A^p}^2+ {q_B^p}^2-Y_p q_A^p q_B^p)}}{ \cosh r_k^p \sqrt{1-e^{4i\phi_k^p} \tanh^2r_k^p}}
\end{equation}
Here $X^{1,2}$ and $Y^{1,2}$ are defined as
\bea
\label{eq18}
X_p=\frac{e^{-4i\phi^{p}_k} \tanh^2 r^{p}_k+1}{2\left(e^{-4i\phi^{p}_k} \tanh^2 r^{p}_k-1\right)};  Y_{p}=\frac{2e^{2i\phi^{p}_k} \tanh r_k^{p}}{e^{-4i\phi^{p}_k} \tanh^2 r^{p}_k-1}. \nno
\eea
Where the \textit{subscript} 'A' stands for momentum mode $\textbf{k}$  and 'B' stands for the momentum mode $-\textbf{k}$. Superscripts $1,2$ are the polarization index. In the case of scalar field also, we can see similar product states of the polarization. However, we can see that the state we obtained depends on the squeezing parameter $r_k$. The evolution of the squeezing parameter depends on the Hamiltonian of the system under consideration. As the evolution of the scalar and gauge fields are totally different, the evolution of the squeezing parameter will also be different. 
Therefore, it is obvious from the present discussion that the large-scale magnetic field and gravitational waves should be in a highly non-classical state at present. 
Quantum discord, an important measure of quantumness, is indeed shown to increase \cite{Martin:2015qta} for such state with increasing $r_k$. However, its measurement in the cosmological setting is not obvious. Therefore,  
in order to quantify this quantumness, we will perform fuzziness of the quantum Poincare sphere on this state Eq.\eqref{state_squeeze} and to show the concreteness of this proposal, we also perform the well established Bell test on this state. For this purpose, we consider two sets of observables with their unique algebraic properties. 
\subsection{Effects of initial non-BD vacuum on the squeezed state}
In the previous section, we have shown the effects of initial excitation on the power spectrum of magnetic fields and primordial gravitational waves. Here we investigate the effects on the squeezed state due to the initial excitation. If we fix the parameters of the function described in Eq.\eqref{beta_mod}, then it fixes the value of $\beta_k$ and thus the excited state $|m_k(\eta_0)\rangle$ also gets fixed, from which the evolution starts. So the final squeezed state will evolve from the state $|m_k^0,m_{-k}^0\rangle$ with a non-zero particle number. As we have taken the factor $g_k$ to be independent of the polarization, this initial state is the same for both polarization modes. In the interaction picture, the state and the operators evolve with time. So the creation and annihilation operators $a_k(0),a_k^{\dagger}(0)$ defined with respect to the BD vacuum also evolves to some $\hat{b}_k$ and $\hat{b}_k^{\dagger}$ depending on the Hamiltonian of the system. And the excited state $|m_k^0, m_{-k}^0\rangle$ acts as a vacuum state with respect to the new creation and annihilation operators $\hat{b}_k,\hat{b}_k^{\dagger}$. So the operator algebra gives
$\hat{b}_k~|m_k^0,m_{-k}^0\rangle=0$. Therefore the new squeezing and displacement operator is defined with respect to the new creation and annihilation operator $\hat{b}_k, \hat{b}_{-k}$ and $\hat{b}_k^{\dagger}, \hat{b}_{-k}^{\dagger}$ respectively as $S'(r'_k,\phi'_k)= e^{B'_k}$ and $R'(\theta'_k)=e^{D'_k}$, where
\bea
B'_k &=& r'_k e^{-2i\phi'_k}\hat{b}_{-k}\hat{b}_k - r'_k e^{2 i\phi'_k} \hat{b}_k^{\dagger}\hat{b}_{-k}^{\dagger}\nno\\
D'_k &=& -i \theta'_k \hat{b}_k^{\dagger}\hat{b}_k -i \theta'_k . \hat{b}_{-k}^{\dagger}\hat{b}_{-k}
\eea 
Here, we have used a different squeezing parameter $r'_k$ as its value starts from $r_0$, unlike the case for general BD vacuum. The final squeezed state will be constructed as
\bea\label{sqzd_state_ex}
|\psi\rangle_{exc}&=& S'_2(r_k^{'2},\phi_k^{'2})R'_2(\theta_k^{'2})|m_k^0,m_{-k}^0\rangle \nno\\
&\otimes& S'_1(r_k^{'1},\phi_k^{'1})R'_1(\theta_k^{'1})|m_k^0,m_{-k}^0\rangle\nno\\
&=& \frac{1}{{\cal A'}}\sum_{n,m=0}^\infty e^{-2i(n\phi_k^{'1} + m\phi_k^{'2})}\tanh^n r_k^{'1}\nno\\
&\cdot& \tanh^m r_k^{'2}\vert n_{k,-k}^1,m_{k,-k}^2 \rangle 
\eea
Where, ${\cal A'} = \cosh (r_k^{'1}) \cosh(r_k^{'2})$. This squeezed state which evolved from the initial excited state, can be projected in the real space as described in Eq.\eqref{state}. It can also be noticed that the state described in Eq.\eqref{sqzd_state_ex} looks exactly similar to the state in Eq.\eqref{state_squeeze}, with the only difference being the squeezing parameter $r$ as in the latter case, it starts to evolve from the value $r_0$.

\section{Quantum Stokes operators}
The Stokes parameters \cite{stokes} are known to be an important description of the polarization properties of the electromagnetic field. The same can be applied to gravitational waves also. 
The quantum Stokes parameters 
are the operator representations of the polarization that can be applied to non-classical waves. Most interestingly these operators satisfy angular momentum like commutation algebra. Stokes operator are defined as follows,
\bea
{\cal S}^{(0)}_k &=& {a_k^1}^{\dagger} a_k^1 + {a_k^2}^{\dagger} a_k^2 ~~;~~ {\cal S}^{(1)}_k = {a_k^1}^{\dagger} a_k^1 - {a_k^2}^{\dagger} a_k^2  \\
{\cal S}^{(2)}_k &=& t_k {a_k^1}^{\dagger} a_k^2 + t^*_k  {a_k^2}^{\dagger} a_k^1 ~~;~~ {\cal S}^{(3)}_k = i(t^*_k {a_k^2}^{\dagger} a_k^1 -t_k {a_k^1}^{\dagger} a_k^2)\nonumber 
\eea
Where $t_k = e^{i\psi_k}$ measures the relative phase between the two modes.
According to the above definition $p=(1,2)$ correspond to two different polarization modes associated with  electromagnetic and gravitational field. $\mathcal{S}_k^{0}$ encodes the intensity of a particular mode $k$ or the total number of photons or graviton. $\mathcal{S}_k^{1}~ \mbox{and}~\mathcal{S}_k^{2}$ measures the linear polarization and $\mathcal{S}_k^{3}$ measures the circular polarization. Interestingly last three Stokes parameters parametrizing the polarization states satisfy $SU(2)$ spin algebra $[{\cal S}^{(p)}_{k}, {\cal S}^{(q)}_{k}] = 2i \epsilon^{pqr} {\cal S}^{(r)}_{k}$, where $\epsilon^{pqr}$ is the three dimensional levicivita tensor density. $(p,q,r)\equiv 1,2,3$. 
In order to prove the quantumness of a system through the Stokes parameters, we examine it via the quantum Poincare sphere\cite{poincare1,poincare2}. For the classical case, we have the usual relation $\langle (\mathcal{S}_k^1)^2+(\mathcal{S}_k^2)^2+(\mathcal{S}_k^3)^2\rangle=\langle(\mathcal{S}_k^0)^2\rangle$\cite{malikyn,jyrki}. But in case of quantum Poincare sphere, we cannot precisely define the radius of the Poincare sphere due to the non-commutativity of $\mathcal{S}^i_k$. And we have $\langle(\mathcal{S}_k^1)^2+(\mathcal{S}_k^2)^2+(\mathcal{S}_k^3)^2-(\mathcal{S}_k^0)^2\rangle> 0$. We define fuzziness of quantum  Poincare sphere (can also be called fuzzy sphere) as a measure of the quantumness
\beq\label{diffusivity}
Q_{fuzzy}=\langle(\mathcal{S}_k^1)^2+(\mathcal{S}_k^2)^2+(\mathcal{S}_k^3)^2-(\mathcal{S}_k^0)^2\rangle
\eeq
In order to calculate the fuzziness of the Poincare sphere constructed by the stokes operators we need to explicitly calculate the expectation values of the operators $(\mathcal{S}_k^i)^2$. We have 
\bea\label{stokes_square}
(\mathcal{S}_k^0)^2 &=&(n_k^1)^2+ n_k^1 n_k^2+ n_k^2 n_k^1+(n_k^2)^2\nno\\
(\mathcal{S}_k^1)^2 &=&(n_k^1)^2- n_k^1 n_k^2- n_k^2 n_k^1+(n_k^2)^2\\
(\mathcal{S}_k^2)^2 &=&(t_k)^2 (a_k^{\dagger 1})^2 (a_k^2)^2+ t_k t_k^* ~n_k^1 a_k^2 a_k^{\dagger 2}+ t_k t_k^* ~n_k^2 a_k^1a_k^{\dagger 1}\nno\\
&+&(t_k^*)^2 (a_k^{\dagger 2})^2 (a_k^1)^2\nno\\
(\mathcal{S}_k^3)^2 &=& - (t_k^*)^2 (a_k^{\dagger 2})^2 (a_k^1)^2 + t_k t_k^* ~n_k^1 a_k^2 a_k^{\dagger 2}+ t_k t_k^* ~n_k^2 a_k^1a_k^{\dagger 1}\nno\\
&-& (t_k)^2 (a_k^{\dagger 1})^2 (a_k^2)^2\nno
\eea
Where $n_k^1=a_k^{\dagger 1} a_k^1$ is the number operator corresponding to polarization 1 and $n_k^2=a_k^{\dagger 2} a_k^2$ is the number operator corresponding to polarization 2. Calculating the expectation values of the operators in Eq.\eqref{stokes_square} on the state defined in Eq.\eqref{state_squeeze} we get
\bea
\langle (\mathcal{S}_k^0)^2\rangle &=& \frac{1}{\mathcal{A}^2}\sum_{m,n=0}^{\infty}\tanh^{2n}(r_k^1)\tanh^{2m}(r_k^2)(n^2+2 mn +m^2)\nno\\
&=&\frac{1}{4}\left(\cosh(4r_k^1)+2 \cosh(2r_k^1)\cosh(2r_k^2)+\cosh(4r_k^2)\right)\nno\\
\langle (\mathcal{S}_k^1)^2\rangle &=&\frac{1}{\mathcal{A}^2}\sum_{m,n=0}^{\infty}\tanh^{2n}(r_k^1)\tanh^{2m}(r_k^2)(n^2-2 mn +m^2)\nno\\
&=& \frac{1}{4}\left(\cosh(4r_k^1)-2 \cosh(2r_k^1)\cosh(2r_k^2)+\cosh(4r_k^2)\right)\nno\\
\langle (\mathcal{S}_k^2)^2\rangle &=&\frac{1}{\mathcal{A}^2}\sum_{m,n=0}^{\infty}\tanh^{2n}(r_k^1)\tanh^{2m}(r_k^2)m(n+1)\nno\\
&=& \cosh^2(r_k^1) \sinh^2(r_k^2)\\
\langle (\mathcal{S}_k^3)^2\rangle &=& \frac{1}{\mathcal{A}^2}\sum_{m,n=0}^{\infty}\tanh^{2n}(r_k^1)\tanh^{2m}(r_k^2)n(m+1)\nno\\
&=&\sinh^2(r_k^1)\cosh^2(r_k^2)\nno
\eea
Where $\mathcal{A}=\cosh(r_k^1)\cosh(r_k^2)$ and $n,m$ are the number operator expectation values on the state for polarization 1 and polarization 2 respectively. Calculating the fuzziness on the state from the expectation values of the operators we get and symbolized as 
\beq\label{poincare_sphere}
Q_{fuzzy}=\langle(\mathcal{S}_k^1)^2+(\mathcal{S}_k^2)^2+(\mathcal{S}_k^3)^2-(\mathcal{S}_k^0)^2\rangle = 4~\sinh^2(r_k)
\eeq
Where for simplicity we have taken $r_k^1=r_k^2=r_k$.
We can see that the quantity is dependent on the squeezing parameter $r_k$. Similarly, when we consider the initial state as an excited state, the final state corresponds to the state mentioned in Eq.\eqref{sqzd_state_ex}. Here to perform the fuzziness of the Poincare sphere, we define the quantum Stokes operators in terms of the new creation and annihilation operators $\hat{b}_k,\hat{b}_{-k}$ and $\hat{b}_k^{\dagger}, \hat{b}_{-k}^{\dagger}$ respectively. Here the Stokes operators are defined as
\bea\label{stokes_ex}
{\cal S'}^{(0)}_k &=& \hat{b}_k^{1\dagger} \hat{b}_k^1 + {\hat{b}_k}^{2\dagger} \hat{b}_k^2 ~~;~~ {\cal S'}^{(1)}_k = {\hat{b}_k}^{\dagger} \hat{b}_k^1 - {\hat{b}_k}^{2\dagger} \hat{b}_k^2  \\
{\cal S'}^{(2)}_k &=& t_k {\hat{b}_k}^{1\dagger} \hat{b}_k^2 + t^*_k  {\hat{b}_k}^{2\dagger} \hat{b}_k^1 ~;~ {\cal S'}^{(3)}_k = i(t^*_k {\hat{b}_k}^{2\dagger} \hat{b}_k^1 -t_k {\hat{b}_k}^{1\dagger} \hat{b}_k^2)\nonumber 
\eea
Now, if the system starts to evolve with some initial excitation, then also we define the fuzziness of the Poincare sphere exactly as defined in Eq.\eqref{diffusivity} with the Stokes operators defined in Eq.\eqref{stokes_ex}. For this system, the fuzziness of the Poincare sphere, having the state defined by Eq.\eqref{sqzd_state_ex} we get,
\beq\label{poincare_sphere_ex}
Q'_{fuzzy}=\langle(\mathcal{S'}_k^1)^2+(\mathcal{S'}_k^2)^2+(\mathcal{S'}_k^3)^2-(\mathcal{S'}_k^0)^2\rangle = 4~\sinh^2(r'_k)\nno
\eeq
Where we have taken $r_k^{'1}=r_k^{'2}=r'_k$. We can see from the results in Eqs.\eqref{poincare_sphere} and \eqref{poincare_sphere_ex} that
when the squeezing parameter $(r_k,r'_k)$ goes to zero, we can consider the state to be classical. Therefore, the higher the squeezing parameter, the higher would be the fuzziness of the Poincare sphere, and thus the quantumness increases.

\section{Measurement and Poincare sphere}
Performing the test for quantumness is not an easy task as for cosmological scenarios. However, the primordial gravitational Eq.\eqref{PGW_power_spectra_present} and magnetic power spectrum  Eq.\eqref{power_spectra} provide us a good observational measure of the squeezing parameters  $(r_k,\phi_k)$, in terms of their strength and index. Using those measured values of $(r_k,\phi_k)$ we can theoretically obtain the fuzziness of the Poincare sphere from \eqref{diffusivity}. Therefore, for a particular value of the squeezing parameter, if that quantity is greater than zero, we can infer about the quantumness of the system. 
 Further, Stokes parameters are also measurable quantities \cite{berry,korolkova}, which are directly related to the fuzziness. From Eq.\eqref{diffusivity} we can see that fuzziness is defined in terms of $\langle(\mathcal{S}_k^i)^2\rangle$ . The operators $(\mathcal{S}_k^{0,1})^2$  are functions of the number operators. Thus it can be measured from the intensity of GW and strength of the magnetic field, respectively. But the expectation values of the operators $(\mathcal{S}_k^{2,3})^2$ can't be measured directly as it contains several correlation functions like as $(a_k^{\dagger i})^2~,~n_k^i a_k^j a_k^{\dagger j}$. However, Agarwal et al. \cite{agarwal} in their paper have shown that these correlation functions can be measured in terms of the intensity and intensity fluctuations. The variance is defined as
 \beq
V_{ij}=\frac{1}{2}\left(\langle\lbrace\mathcal{S}_i,\mathcal{S}_j\rbrace\rangle-\lbrace\langle\mathcal{S}_i\rangle,\langle\mathcal{S}_j\rangle\rbrace \right) 
 \eeq
Where the curly brace is the anti-commutator. The diagonal elements can be calculated as
\beq
V_{ii}=\langle(\mathcal{S}_i)^2\rangle-(\langle\mathcal{S}_i\rangle)^2
\eeq
As this quantity $V_{ii}$ can be measured \cite{korolkova,mukunda} and the expectation values of the Stokes parameter $\langle\mathcal{S}_k^i\rangle$  is a measurable quantity we can write Eq.\eqref{diffusivity} as
\beq
\label{diff_measure}
Q_{fuzzy}=\langle (\mathcal{S}_k^1)^2+ V_{22} - (\langle\mathcal{S}_k^2\rangle)^2 + V_{33} - (\langle\mathcal{S}_k^3\rangle)^2-(\mathcal{S}_k^0)^2\rangle
\eeq

Therefore, experimentally measured value of the fuzziness $\langle(S_k^1)^2+(S_k^2)^2+(S_k^3)^2-(S_k^0)^2\rangle$ in terms of stokes parameters $(\mathcal{S}_k^1, \mathcal{S}_k^2,\mathcal{S}_k^3)$, can now be compared with its classical counterpart (which is zero)\cite{malikyn,jyrki}. Upon observing $\langle(\mathcal{S}_k^1)^2+(\mathcal{S}_k^2)^2+(\mathcal{S}_k^3)^2-(\mathcal{S}_k^0)^2\rangle > 0$, we can infer the quantum origin of those primordial fluctuations. In principle when we make the measurement of either gravitational wave or primordial magnetic field, we cannot measure a particular wave number $k$. We measure a superposition of several wave numbers. Therefore the actual measurable quantity is defined as
\beq\label{sum_diff}
\sum_{k=k_{min}}^{k_{max}} \langle(\mathcal{S}_k^1)^2+(\mathcal{S}_k^2)^2+(\mathcal{S}_k^3)^2-(\mathcal{S}_k^0)^2\rangle
\eeq
We know for classical case the quantity defined in Eq.\eqref{diffusivity} $Q_{fuzzy}$ is zero, so it is never negative for any scenario. Thus the measurement will give us
\beq
\sum_{k=k_{min}}^{k_{max}} \langle(\mathcal{S}_k^1)^2+(\mathcal{S}_k^2)^2+(\mathcal{S}_k^3)^2-(\mathcal{S}_k^0)^2\rangle \geq 0
\eeq
After the measurement, if the measured quantity is greater than zero, we can conclude that the system under consideration has a quantum origin.


\section{Pseudospin operators and Bell violation}
Now we embark on a different set of operators called pseudo spin operators and their Bell violation test to further support our proposal in the last section. This set of operators can be very useful in several quantum mechanical systems. We will discuss about the application of those in future works. When dealing with the system having continuous variables, there exists an elegant way \cite{Martin:2017zxs,larsson,gour} of constructing a set of position-dependent pseudo spin operators which satisfy the same algebra as the spin angular momentum. {\em To the best of our knowledge, such operators were constructed only in one dimension}. Our present study demands the generalization of those operators in two dimensions corresponding to two independent polarization modes. 
As a prerequisites of the spin like operators following operators are defined in position $(q,q')$ space, 
\begin{eqnarray}
\label{eq21}
P_{n,m}(l) &=&\int_{nl}^{(n+1)l} dq \int_{ml}^{(m+1)l} dq'\sss|q,q'\rangle\langle q,q'|, \\
T_{n,m}(l)&=&\int_{nl}^{(n+1)l} dq \int_{ml}^{(m+1)l} dq'\sss|q,q'\rangle\langle q+l,q'+l|,\nno\\
T^\dagger_{n,m}(l)&=&\int_{(n+1)l}^{(n+2)l} dq \int_{(m+1)l}^{(m+2)l} dq'\sss|q,q'\rangle\langle q-l,q'-l|.\nno
\end{eqnarray}
Those are called on-site projection, forward and backward hopping operators respectively. When the projection and the hopping operators operates on a state they behave as
\beq
P_{n,m}(l)~\psi(q)=\left\{\begin{array}{ll}
\psi(q) &~~ml,nl\leq q<ml+l,nl+l\\
0 & ~~\mbox{otherwise}
\end{array}\right.
\eeq
\beq
T_{n,m}(l)~\psi(q)=\left\{\begin{array}{ll}
\psi(q+l) &~~ml,nl\leq q<ml+l,nl+l\\
0 & ~~\mbox{otherwise}
\end{array}\right.
\eeq
\beq
T_{n,m}^{\dagger}(l)~\psi(q)=\left\{\begin{array}{ll}
\psi(q-l) &ml+l,nl+l\leq q<ml+2l,nl+2l\\
0 & \mbox{otherwise}
\end{array}\right.
\eeq
The generalized pseudo spin operators are constructed from the aforementioned position space operators and generalizing from \cite{larsson} we define them as
\begin{eqnarray}
\label{eq22}
s_z &=&\sum_{{\cal  C}=-\infty}^\infty(-1)^{n+m} P_{n,m}\nno\\
s_+&=&\sum_{{\cal  C}=-\infty}^\infty T_{2n,2m} \\
s_- &=& \sum_{{\cal  C}=-\infty}^\infty T_{2n,2m} ^{\dagger}\nno 
\end{eqnarray}
where, ${\cal C} \equiv (m,n)$, and $l$ is the length of the interval. We define $s_+=s_x+i~ s_y $ and $s_-=s_x-i~ s_y$. It can be easily shown that these three operators $s_x,s_y$ and $s_z$ are dichotomous i.e ($s_x^2=s_y^2=s_z^2=1$). And they follow the usual $SU(2)$ spin algebra.  
Exploiting the nature of the operators we can define the Bell-CHSH operator in terms of the pseudospin operators. For two observer 'A' having momentum $\textbf{k}$ and  observer 'B' having momentum $\textbf{-k}$, the maximal Bell-CHSH operator can be constructed as
\beq
\label{chsh_pseudo}
B^s_{CHSH}\equiv 2\sqrt{\left(s_z^A\otimes s_z^B\right)^2+\left(s_x^A\otimes s_x^B\right)^2 }
\eeq
On the quantum state Eq.\eqref{state} we now calculate the expectation value of the Bell operator $B^s_{CHSH}$  with $r_k^1=r_k^2=r$ presented in
\begin{figure}
\includegraphics[scale=0.9]{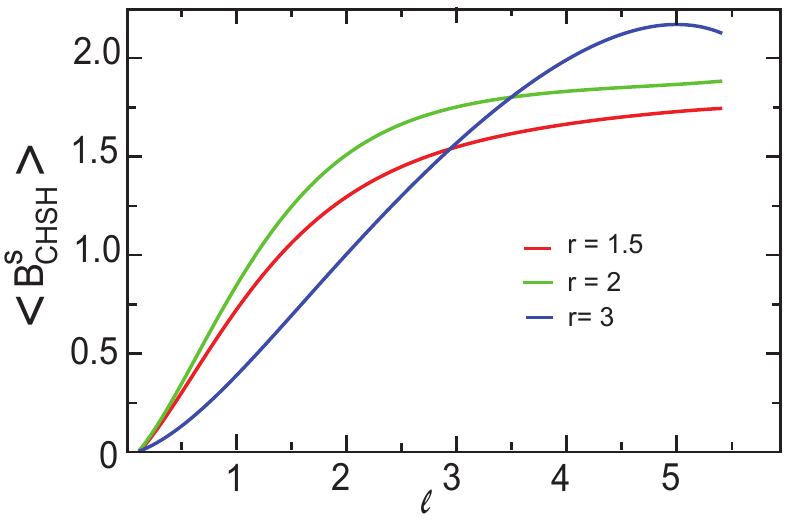}
\begin{center}
\caption{Variation of the expectation value of $B^s_{CHSH}$ with the interval length $l$ for different squeezing parameters $r_k$ for $\theta_k-\frac{\phi_k}{2}=0$} \label{pseudo}
\end{center}
\end{figure}
the Fig.\ref{pseudo}. We know that the Bell inequality suggests that for a system having the Bell-CHSH operator expectation value grater than 2 can only be described by means of quantum mechanics. Thus it is evident from the figure that the system behaves quantum mechanically, i.e $\langle B^s_{CHSH}\rangle > 2$,  above a certain value of the squeezing parameter $r_k$, which is in agreement with conclusion from our previous Poincare sphere measurement.
\section{Conclusion}
In this letter, we have proposed a Poincare sphere fuzziness or Fuzzy sphere in terms of observable quantities to verify the quantum origin of the large-scale magnetic field and gravitational waves. For this purpose, we first constructed the squeezed quantum states for the electromagnetic and gravitational wave fluctuation during inflation and the associated power spectrum. We have seen that the evolution of the squeezing parameter is different for the case of primordial magnetic fields and GW. In case of primordial magnetic field asymptotically gives the values $r_k$ = 6.9, 14.9 and 23.4 for the coupling parameter $n$ = 1, 2 and 3 respectively. In the case of GW, the squeezing parameter $r_k$ increases asymptotically. So we can expect information about the quantum origin of the fields. 
Furthermore, if we consider an initial non-BD vacuum, the initial squeezing parameter naturally becomes non-zero. For example, if we assume $r_0=0.51$ for a mode $k=0.8 $MPc$^{-1}$, the final values of the squeezing parameter turn out as $r_{k=0.8} = (9.468, 9.550, 9.621)$ for $\gamma =(3,4,5)$ respectively in case of primordial magnetic field for the coupling parameter $n=1$. Therefore, due to  an initial excitation, there is an increase in the final value of the squeezing parameter at the end of inflation. We have then considered the Poincare sphere fuzziness or the uncertainty of the radius in terms of quantum stokes operators, which assume a positive value for the cosmologically evolved squeezed quantum states. Although all the terms described in Eq.\eqref{diffusivity} cannot be measured directly, we can measure all the terms of the variances and intensities of the Stokes parameters as defined in Eq.\eqref{diff_measure}. 
Our theoretical calculation reveals that the larger the squeezing, the stronger the quantumness. Importantly this can be confirmed by the direct measurement of Stokes parameters and primordial power spectrums. Finally, we confirmed our conclusion by further performing the Bell test considering dichotomic pseudospin operators. Other quantum diagnostics such as quantum discord and Fano factors could be interesting to study in this present context.


\begin{thebibliography}{99}
\bibitem{Starobinsky:1980te}
A.~A.~Starobinsky,
Adv. Ser. Astrophys. Cosmol. \textbf{3}, 130-133 (1987)
\bibitem{Starobinsky:1982ee}
A.~A.~Starobinsky,
Phys. Lett. B \textbf{117}, 175-178 (1982)
\bibitem{Guth:1980zm}
A.~H.~Guth,
Phys. Rev. D \textbf{23}, 347-356 (1981)
\bibitem{Linde:1981mu}
A.~D.~Linde,
Phys. Lett. B \textbf{108}, 389-393 (1982)
\bibitem{Albrecht:1982wi}
A.~Albrecht and P.~J.~Steinhardt,
Phys. Rev. Lett. \textbf{48}, 1220-1223 (1982)
\bibitem{Linde:1983gd}
A.~D.~Linde,
Phys. Lett. B \textbf{129}, 177-181 (1983)
\bibitem{Martin:2012pea}
J.~Martin, V.~Vennin and P.~Peter,
Phys. Rev. D \textbf{86}, 103524 (2012)
[arXiv:1207.2086 [hep-th]].
\bibitem{Martin:2015qta}
J.~Martin and V.~Vennin,
Phys. Rev. D \textbf{93}, no.2, 023505 (2016)
[arXiv:1510.04038 [astro-ph.CO]].
\bibitem{Martin:2016nrr}
J.~Martin and V.~Vennin,
Phys. Rev. A \textbf{94}, no.5, 052135 (2016)
[arXiv:1611.01785 [quant-ph]].
\bibitem{Martin:2016tbd}
J.~Martin and V.~Vennin,
Phys. Rev. A \textbf{93}, no.6, 062117 (2016)
[arXiv:1605.02944 [quant-ph]].
\bibitem{Martin:2017zxs}
J.~Martin and V.~Vennin,
Phys. Rev. D \textbf{96}, no.6, 063501 (2017)
[arXiv:1706.05001 [astro-ph.CO]].
\bibitem{Kanno:2019gqw}
S.~Kanno,
Phys. Rev. D \textbf{100}, no.12, 123536 (2019)
[arXiv:1905.06800 [hep-th]].
\bibitem{Sharma:2017eps}
R.~Sharma, S.~Jagannathan, T.~R.~Seshadri and K.~Subramanian,
Phys. Rev. D \textbf{96} (2017) no.8, 083511
[arXiv:1708.08119 [astro-ph.CO]].

\bibitem{Sharma:2018kgs}
R.~Sharma, K.~Subramanian and T.~R.~Seshadri,
Phys. Rev. D \textbf{97} (2018) no.8, 083503
[arXiv:1802.04847 [astro-ph.CO]].


\bibitem{Jain:2012ga}
R.~K.~Jain and M.~S.~Sloth,
Phys. Rev. D \textbf{86} (2012), 123528
[arXiv:1207.4187 [astro-ph.CO]].



\bibitem{Durrer:2010mq}
R.~Durrer, L.~Hollenstein and R.~K.~Jain,
JCAP \textbf{03} (2011), 037
[arXiv:1005.5322 [astro-ph.CO]].


\bibitem{Kanno:2009ei}
S.~Kanno, J.~Soda and M.~a.~Watanabe,
JCAP \textbf{12} (2009), 009
[arXiv:0908.3509 [astro-ph.CO]].


\bibitem{Campanelli:2008kh}
L.~Campanelli,
Int. J. Mod. Phys. D \textbf{18} (2009), 1395-1411
[arXiv:0805.0575 [astro-ph]].




\bibitem{Ashoorioon:2004rs}
A.~Ashoorioon and R.~B.~Mann,
Phys. Rev. D \textbf{71}, 103509 (2005)
[arXiv:gr-qc/0410053 [gr-qc]].
\bibitem{Demozzi:2009fu}
V.~Demozzi, V.~Mukhanov and H.~Rubinstein,
JCAP \textbf{08} (2009), 025
[arXiv:0907.1030 [astro-ph.CO]].
\bibitem{Abbott:2016blz}
B.~P.~Abbott \textit{et al.} [LIGO Scientific and Virgo],
Phys. Rev. Lett. \textbf{116}, no.6, 061102 (2016)
[arXiv:1602.03837 [gr-qc]]; 
\bibitem{lv2} C. Caprini and D. G. Figueroa, Class. Quant. Grav. {\bf 35}, 163001 (2018), arXiv:1801.04268 [astro-ph.CO];
\bibitem{lv3} B. Abbott et al. (LIGO Scientific, Virgo), Phys. Rev. Lett. {\bf 116}, 131103 (2016), arXiv:1602.03838 [gr-qc]; 
\bibitem{lv4} B. Abbott et al. (LIGO Scientific, Virgo), Phys. Rev. D {\bf 93}, 122003 (2016), arXiv:1602.03839 [gr-qc]; 
\bibitem{lv5} B. Abbott et al. (LIGO Scientific, Virgo), Phys. Rev. Lett. {\bf 116}, 241102 (2016), arXiv:1602.03840 [gr-qc]; 
\bibitem{lv6} B. P. Abbott et al. (LIGO Scientific, Virgo), Phys. Rev. Lett. {\bf 116}, 061102 (2016), arXiv:1602.03837 [gr-qc]; 
\bibitem{lv7} B. P. Abbott et al. (LIGO Scientific, Virgo), Phys. Rev. Lett. {\bf 116}, 241103 (2016), arXiv:1606.04855 [gr-qc]; 
\bibitem{lv8} B. P. Abbott et al. (LIGO Scientific, VIRGO), Phys. Rev. Lett. {\bf 118}, 221101 (2017), [Erratum: Phys. Rev.
Lett. {\bf 121},no.12,129901(2018)], arXiv:1706.01812 [gr-qc];
\bibitem{lv9} B. P. Abbott et al. (LIGO Scientific, Virgo), Astrophys. J. Lett. {\bf 851}, L35 (2017), arXiv:1711.05578 [astro-ph.HE];
\bibitem{lv10} B. P. Abbott et al. (LIGO Scientific, Virgo), Phys. Rev. Lett. {\bf 119}, 141101 (2017), arXiv:1709.09660 [gr-qc]; 
\bibitem{lv11} B. P. Abbott et al. (LIGO Scientific, Virgo), Phys. Rev. Lett. {\bf 119}, 161101 (2017), arXiv:1710.05832 [gr-qc]; 
\bibitem{lv12} R. Abbott et al. (LIGO Scientific, Virgo), Phys. Rev. D {\bf 102}, 043015 (2020), arXiv:2004.08342 [astro-ph.HE]; 
\bibitem{lv13} B. P. Abbott et al. (LIGO Scientific, Virgo), Astrophys. J. Lett. {\bf 892}, L3 (2020), arXiv:2001.01761 [astro-ph.HE]; 
\bibitem{lv14} R. Abbott et al. (LIGO Scientific, Virgo), Astrophys. J. {\bf 896}, L44 (2020), arXiv:2006.12611 [astro-ph.HE].

\bibitem{Grasso:2000wj}
D.~Grasso and H.~R.~Rubinstein,
Phys. Rept. \textbf{348}, 163-266 (2001)
[arXiv:astro-ph/0009061 [astro-ph]].
\bibitem{Kronberg:1993vk}
P.~P.~Kronberg,
Rept. Prog. Phys. \textbf{57}, 325-382 (1994)
\bibitem{Widrow:2002ud}
L.~M.~Widrow,
Rev. Mod. Phys. \textbf{74}, 775-823 (2002)
[arXiv:astro-ph/0207240 [astro-ph]].
\bibitem{Durrer:2013pga}
R.~Durrer and A.~Neronov,
Astron. Astrophys. Rev. \textbf{21}, 62 (2013)
[arXiv:1303.7121 [astro-ph.CO]].
\bibitem{Kobayashi:2014sga}
T.~Kobayashi,
JCAP \textbf{05}, 040 (2014)
[arXiv:1403.5168 [astro-ph.CO]].
\bibitem{Kobayashi:2019uqs}
T.~Kobayashi and M.~S.~Sloth,
Phys. Rev. D \textbf{100}, no.2, 023524 (2019)
[arXiv:1903.02561 [astro-ph.CO]].
\bibitem{Ferreira:2013sqa}
R.~J.~Z.~Ferreira, R.~K.~Jain and M.~S.~Sloth,
JCAP \textbf{10}, 004 (2013)
[arXiv:1305.7151 [astro-ph.CO]].
\bibitem{Martin:2007ue}
J.~Martin and J.~Yokoyama,
JCAP \textbf{01}, 025 (2008)
[arXiv:0711.4307 [astro-ph]].
\bibitem{dolginov}
A.Z. Dolginov, Phys. Rep. 162 (1988) 337.



\bibitem{asseo}
E. Asseo and H. Sol, Phys. Rep. 148 (1987) 307.
\bibitem{Ratra:1991bn}
B.~Ratra,
Astrophys. J. Lett. \textbf{391}, L1-L4 (1992)
\bibitem{Vachaspati:2016xji}
T.~Vachaspati,
Phys. Rev. D \textbf{95}, no.6, 063505 (2017)
[arXiv:1606.06186 [astro-ph.CO]].
\bibitem{Kanno:2020lpi}
S.~Kanno and J.~Soda,
Symmetry \textbf{12}, no.4, 672 (2020)
[arXiv:2003.14073 [gr-qc]].

\bibitem{Koh:2004ez}
S.~Koh, S.~P.~Kim and D.~J.~Song,
JHEP \textbf{12}, 060 (2004)
[arXiv:gr-qc/0402065 [gr-qc]].
\bibitem{stokes}
G. G. Stokes, Trans. Camb. Phil. Soc. 9, 399 (1852).
\bibitem{poincare1}
Padgett, Miles J., and Johannes Courtial. "Poincaré-sphere equivalent for light beams containing orbital angular momentum." Optics letters 24.7 (1999): 430-432.
\bibitem{poincare2}
Milione, Giovanni, et al. "Higher-order Poincaré sphere, Stokes parameters, and the angular momentum of light." Physical review letters 107.5 (2011): 053601.
\bibitem{malikyn}
Malykin, G. B. "Use of the poincare sphere in polarization optics and classical and quantum mechanics. review." Radiophysics and quantum electronics 40.3 (1997): 175-195. 
\bibitem{jyrki}
Laatikainen, Jyrki, et al. "Poincaré sphere of electromagnetic spatial coherence." Optics Letters 46.9 (2021): 2143-2146.
\bibitem{berry}
Berry, H. Gordon, G. Gabrielse, and A. E. Livingston. "Measurement of the Stokes parameters of light." Applied optics 16.12 (1977): 3200-3205.
\bibitem{korolkova}
Korolkova, Natalia, et al. "Polarization squeezing and continuous-variable polarization entanglement." Physical Review A 65.5 (2002): 052306.
\bibitem{larsson}
Jan-\. Ake Larsson, Physical Review A \textbf{70},022102  (2004)
\bibitem{gour}
G. Gour, F.C. Khanna, A. Mann, and M. Rezven
Physics Letters A \textbf{324}, 415  (2004)
\bibitem{agarwal}
G. S. Agarwal, S. Chaturvedi (2003) Scheme to measure quantum stokes parameters and their fluctuations and correlations, Journal of Modern Optics, 50:5, 711-716, DOI 10.1080/09500340308235179
\bibitem{mukunda}
N. Mukunda and T. F. Jordan, J. Math. Phys. 7, 849 (1966).
\bibitem{Ragavendra:2020vud}
H.~V.~Ragavendra, L.~Sriramkumar and J.~Silk,
JCAP \textbf{05}, 010 (2021)
[arXiv:2011.09938 [astro-ph.CO]].
\bibitem{Holman:2007na}
R.~Holman and A.~J.~Tolley,
JCAP \textbf{05}, 001 (2008)
[arXiv:0710.1302 [hep-th]].
\bibitem{Ashoorioon:2013eia}
A.~Ashoorioon, K.~Dimopoulos, M.~M.~Sheikh-Jabbari and G.~Shiu,
JCAP \textbf{02}, 025 (2014)
[arXiv:1306.4914 [hep-th]].
\bibitem{Ashoorioon:2010xg}
A.~Ashoorioon and G.~Shiu,
JCAP \textbf{03}, 025 (2011)
[arXiv:1012.3392 [astro-ph.CO]].
\bibitem{Meerburg:2009ys}
P.~D.~Meerburg, J.~P.~van der Schaar and P.~S.~Corasaniti,
JCAP \textbf{05}, 018 (2009)
[arXiv:0901.4044 [hep-th]].
\bibitem{Meerburg:2009fi}
P.~D.~Meerburg, J.~P.~van der Schaar and M.~G.~Jackson,
JCAP \textbf{02}, 001 (2010)
[arXiv:0910.4986 [hep-th]].
\bibitem{Ashoorioon:2012kh}
A.~Ashoorioon, P.~S.~Bhupal Dev and A.~Mazumdar,
Mod. Phys. Lett. A \textbf{29}, no.30, 1450163 (2014)
[arXiv:1211.4678 [hep-th]].

\end{thebibliography}
\end{document}